\newcommand{\myname}{Luis Guillermo Natera Orozco}
\newcommand{\myemail}{Corresponding author: misz@itu.dk}
\newcommand{\myaffiliation}{Department of Network and Data Science\\Central European University, H-1051 Budapest, Hungary}
\newcommand{\paperdate}{\today}
\newcommand{\papertitle}{Extracting the multimodal fingerprint of urban transportation networks}
\newcommand{\paperkeywords}{multiplex networks, sustainable transport, multi-modal transportation}

\RequirePackage[l2tabu,orthodox]{nag}   
\documentclass[11pt,onecolumn]{article} 

\usepackage[T1]{fontenc}                
\usepackage[utf8]{inputenc}             
\usepackage{crimson}                    
\usepackage{helvet}                     

\usepackage[strict,autostyle]{csquotes} 
\usepackage[USenglish]{babel}           
\usepackage{microtype}                  

\usepackage{abstract}                   
\usepackage{authblk}                    
\usepackage{booktabs}                   
\usepackage{caption}                    
\usepackage[final]{draftwatermark}      
\usepackage{geometry}                   
\usepackage{graphicx}                   
\usepackage{hyperref}                   
\usepackage{rotating}                   
\usepackage{setspace}                   
\usepackage{titlesec}                   
\usepackage{url}                        
\usepackage[authoryear]{natbib}    
\usepackage{adjustbox}
\usepackage{multirow}
\usepackage{subfigure}

\usepackage{amsmath}
\usepackage{algorithm}
\usepackage[noend]{algpseudocode}

\usepackage{color}

\makeatletter
\def\BState{\State\hskip-\ALG@thistlm}
\makeatother


\titleformat{\section}{\normalfont\sffamily\large\bfseries\color{black}}{\thesection.}{0.3em}{}
\titleformat{\subsection}{\normalfont\sffamily\small\bfseries\color{black}}{\thesubsection.}{0.3em}{}

\hypersetup{
	pdfauthor={\myname},
	pdftitle={\papertitle},
	pdfsubject={\papertitle},
	pdfkeywords={\paperkeywords},
	pdffitwindow=true,         
	breaklinks=true,           
	colorlinks=false,          
	pdfborder={0 0 0}          
}

\SetWatermarkText{DRAFT}
\SetWatermarkScale{1.3}
\SetWatermarkLightness{0.9}

\begin{document}

\title{\papertitle}
\date{\paperdate}
\author{\myname$^1$, Federico Battiston$^{1}$, Gerardo I\~niguez$^{1,2,3}$, Michael Szell$^{4,5,6}$\footnote{\myemail}}

\affil[]{\small
					$1$) \myaffiliation\\
					$2$) Department of Computer Science, Aalto University School of Science, FI-00076 Aalto, Finland\\
					$3$) IIMAS, Universidad Nacional Aut\'{o}noma de M\'{e}xico, 01000 Ciudad de M\'{e}xico, Mexico\\
					$4$) NEtwoRks, Data, and Society (NERDS), IT University of Copenhagen, DK-2300 Copenhagen, Denmark \\
					$5$) ISI Foundation, 10126 Turin, Italy\\
					$6$) Complexity Science Hub Vienna, 1080 Vienna, Austria\\
				}

\maketitle
\begin{onecolabstract}
	Urban mobility increasingly relies on multimodality, combining the use of bicycle paths, streets, and rail networks. These different modes of transportation are well described by multiplex networks. Here we propose the overlap census method which extracts a multimodal profile from a city's multiplex transportation network. We apply this method to 15 cities, identify clusters of cities with similar profiles, and link this feature to the level of sustainable mobility of each cluster. Our work highlights the importance of evaluating all the transportation systems of a city together to adequately identify and compare its potential for sustainable, multimodal mobility.


\vspace{5.5cm}
\end{onecolabstract}

\pagebreak
\section{RESEARCH QUESTION AND HYPOTHESIS}
The infrastructure of different modes of transportation can be described as a mathematical object, the multiplex transport network \citep{Morris2012,Strano2012,Barthelemy2013,Battiston2014a,Gallotti2014,DeDomenico2014,Strano2015,Aleta2017,Lee2017}. A city's multiplex transport network contains the layer of streets and other coevolving network layers, such as the bicycle or the rail networks, which together constitute the multimodal transportation backbone of a city. Due to the car-centric development of most cities \citep{Jacobs1961}, streets form the most developed layers \citep{Gossling2016,Szell2018} and define or strongly limit other layers: For example, sidewalks are by definition footpaths along the side of a street and make up a substantial part of a city's pedestrian space \citep{Gossling2016}. Similarly, most bicycle paths are part of a street or are built along the side. Yet, the different layers of a multimodal network typically serve as diverse channels to permeate a city. Here we consider the transport networks of 15 world cities and develop an urban fingerprinting technique based on multiplex network theory to characterize the various ways in which transport layers can be interconnected, identifying the potential for multimodal transport. Using clustering algorithms on the resulting urban fingerprints, we find distinct classes of cities, reflecting their transport priorities.

\section{METHODS AND DATA}
\begin{figure*}[t!]
	\centering
	\includegraphics[width=\textwidth]{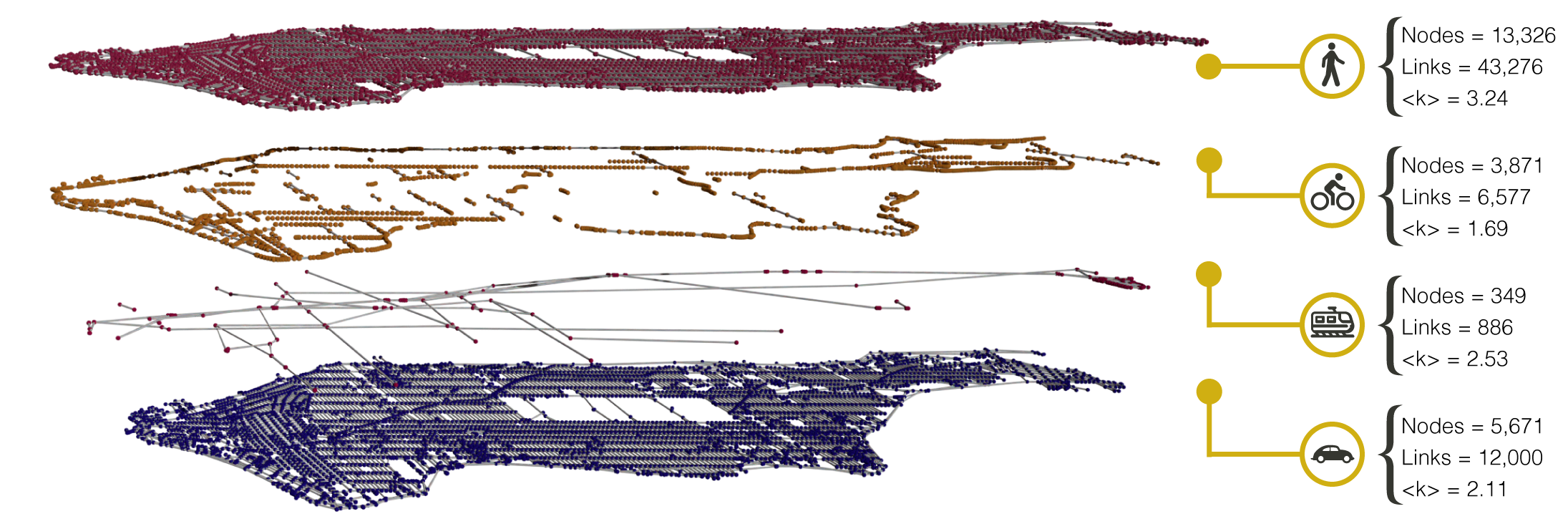}
	\caption{\textbf{(Map plot left)} Multiplex network representation of Manhattan with the four analyzed layers of transport infrastructure (pedestrian paths, bicycle paths, rail lines, and streets), with data from OpenStreetMap. \textbf{(Right)} Network information for each layer, number of nodes, links and average degree $\langle k \rangle$.}
	\label{fig:fig01}
\end{figure*}

We acquired urban transportation networks from multiple cities around the world, defined by their administrative boundaries, using OSMnx \citep{Boeing2017a}. These data sets are of high quality \citep{Haklay2010b,Girres2010} in terms of correspondence with municipal open data \citep{Ferster2019} and completeness \citep{Barbosa-Filho2017}. The various analyzed urban areas and their properties are reported in Table~\ref{tab:Table1}. Figure \ref{fig:fig01} shows the different network layers for Manhattan, one  of our analyzed cities. 

\begin{table*}[ht!]
	\centering
	\begin{adjustbox}{width=\textwidth,keepaspectratio}
		\begin{tabular}{l|rrr|rrr|rrr|rrr|r}
			{} & \multicolumn{3}{c|}{Pedestrian} & \multicolumn{3}{c|}{Bicycle} & \multicolumn{3}{c|}{Rail} & \multicolumn{3}{c|}{Street} &  Population \\
			{} &        Nodes &        Links & $\langle k \rangle$ &       Nodes &       Links & $\langle k \rangle$ &      Nodes &      Links & $\langle k \rangle$ &        Nodes &        Links & {$\langle k \rangle$} \\
			\midrule
			Amsterdam  &   23,321 &  33,665 &  2.89 &  34,529 &  35,619 &  2.06 &  1,096 &  1,655 &  3.02 &   15,125 &  21,722 &  2.87 &     872,680 \\
			Barcelona  &   20,203 &  30,267 &  3.00 &  7,553 &  7,647 &  2.02 &    249 &   249 &  2.00 &  10,393 &  15,809 &  3.04 &   1,600,000 \\
			Beihai     &    2,026 &   2,978 &  2.94 &       0 &      0 &  0.00 &    59 &    62 &  2.10 &    2,192 &   3,209 &  2.93 &   1,539,300 \\
			Bogota     &   81,814 &  121,038 &  2.96 &   9,760 &  9,651 &  1.98 &    166 &   165 &  1.99 &   62,017 &  91,197 &  2.94 &   7,412,566 \\
			Budapest   &   73,172 &  106,167 &  2.90 &  10,494 &  10,318 &  1.97 &  1,588 &  1,964 &  2.47 &   37,012 &  52,361 &  2.83 &   1,752,286 \\
			Copenhagen &   30,746 &  41,916 &  2.73 &  13,980 &  13,988 &  2.00 &   276 &   369 &  2.67 &   15,822 &  20,451 &  2.59 &   2,557,737 \\
			Detroit    &   47,828 &  78,391 &  3.28 &   3,663 &  3,626 &  1.98 &     20 &    21 &  2.10 &   28,462 &  45,979 &  3.23 &     672,662 \\
			Jakarta    &  140,042 &  191,268 &  2.73 &     248 &    231 &  1.86 &     58 &    54 &  1.86 &  138,388 &  188,637 &  2.73 &  10,075,310 \\
			LA         &   89,543 &  128,757 &  2.88 &  14,577 &  14,428 &  1.98 &    173 &   221 &  2.55 &   71,091 &  101,692 &  2.86 &   3,792,621 \\
			London     &  270,659 &  351,824 &  2.60 &  62,398 &  60,043 &  1.92 &  2,988 &  3,535 &  2.37 &  179,782 &  219,917 &  2.45 &   8,908,081 \\
			Manhattan  &   13,326 &  21,447 &  3.22 &   3,871 &  3,777 &  1.95 &    349 &   436 &  2.50 &    5,671 &   9,379 &  3.31 &   1,628,701 \\
			Mexico     &  108,033 &  158,425 &  2.93 &   5,218 &  5,278 &  2.02 &    370 &   364 &  1.97 &   95,375 &  140,684 &  2.95 &   8,918,653 \\
			Phoenix    &  111,363 &  157,075 &  2.82 &  35,631 &  35,979 &  2.02 &    105 &   138 &  2.63 &   73,688 &  102,139 &  2.77 &   1,445,632 \\
			Portland   &   50,878 &  72,958 &  2.87 &  24,252 &  24,325 &  2.01 &    230 &   340 &  2.96 &   35,025 &  49,062 &  2.80 &     583,776 \\
			Singapore  &   82,808 &  110,612 &  2.67 &  12,981 &  12,947 &  1.99 &    683 &   740 &  2.17 &   50,403 &  66,779 &  2.65 &   5,638,700
			\end{tabular}
	\end{adjustbox}
	\caption{Measures for the administrative area of analyzed cities. The number of nodes, links and average degree ($\langle k \rangle$) for each layer in all cities of our dataset are highly diverse due to the varying developmental levels and focus of transport. The range of population in the analyzed cities goes from half million people to ten million people living in Jakarta, this allows to
	have a range of different sizes and cover different developmental stages.  
		\label{tab:Table1}}
\end{table*}

Data and code to replicate the results are available in: (\url{https://doi.org/10.7910/DVN/GSOPCK}), and: (\url{https://github.com/nateraluis/Multimodal-Fingerprint}).

We characterize each city as a multiplex network \citep{Boccaletti2014,Kivela2014,Battiston2017a} with $M$ layers and $N$ nodes that can be active in one or more layers in the system. Layers follow a primal approach \citep{Porta2006} where nodes represent intersections (that may be present in one or more layers), and links represent streets (denoted by \textit{s}), bicycle paths and designated bicycle infrastructure (\textit{b}), subways, trams and rail infrastructure (\textit{r}), or pedestrian infrastructure (\textit{p}). Construction of these intersection nodes follows the topological simplification rules of OSMnx \citep{Boeing2017a}.

In a multimodal city, we expect to find many transport hubs that connect different layers, such as train stations with bicycle and street access, i.e. nodes that are active in different multiplex configurations. Here we propose a method to assess all such combinations of node activities in the system, helping us to learn how well connected different modes are. For each city, we build a profile based on the combinations of node activities, and refer to it as \emph{overlap census} (Figure \ref{fig:fig02}). The overlap census captures the percentage of nodes that are active in different multiplex configurations and provides an ``urban fingerprint'' of its multimodality \citep{Aleta2017}. To define the overlap census formally, given a multiplex transport network with $M$ layers the overlap census is a vector of $(2^M)-1$ components, which accounts for the fractions of nodes that can be reached through at least one layer.

In Fig.~\ref{fig:fig02}(a) we show a schematic of how the overlap census is built: taking the multiplex network, and calculating the percentage of nodes that overlap in different configurations. The multiplex approach addresses the multimodality of a city: it not only counts how many nodes or links there are in each layer, but it shows how they are combined, revealing the possible multimodal mobility combinations in the city. Understanding the possibilities for interchange between mobility layers provides us with a better understanding of urban systems, showing us the complexity and interplay between layers.

\section{FINDINGS}

\begin{figure*}[t!]
	\centering
	\includegraphics[width=\textwidth]{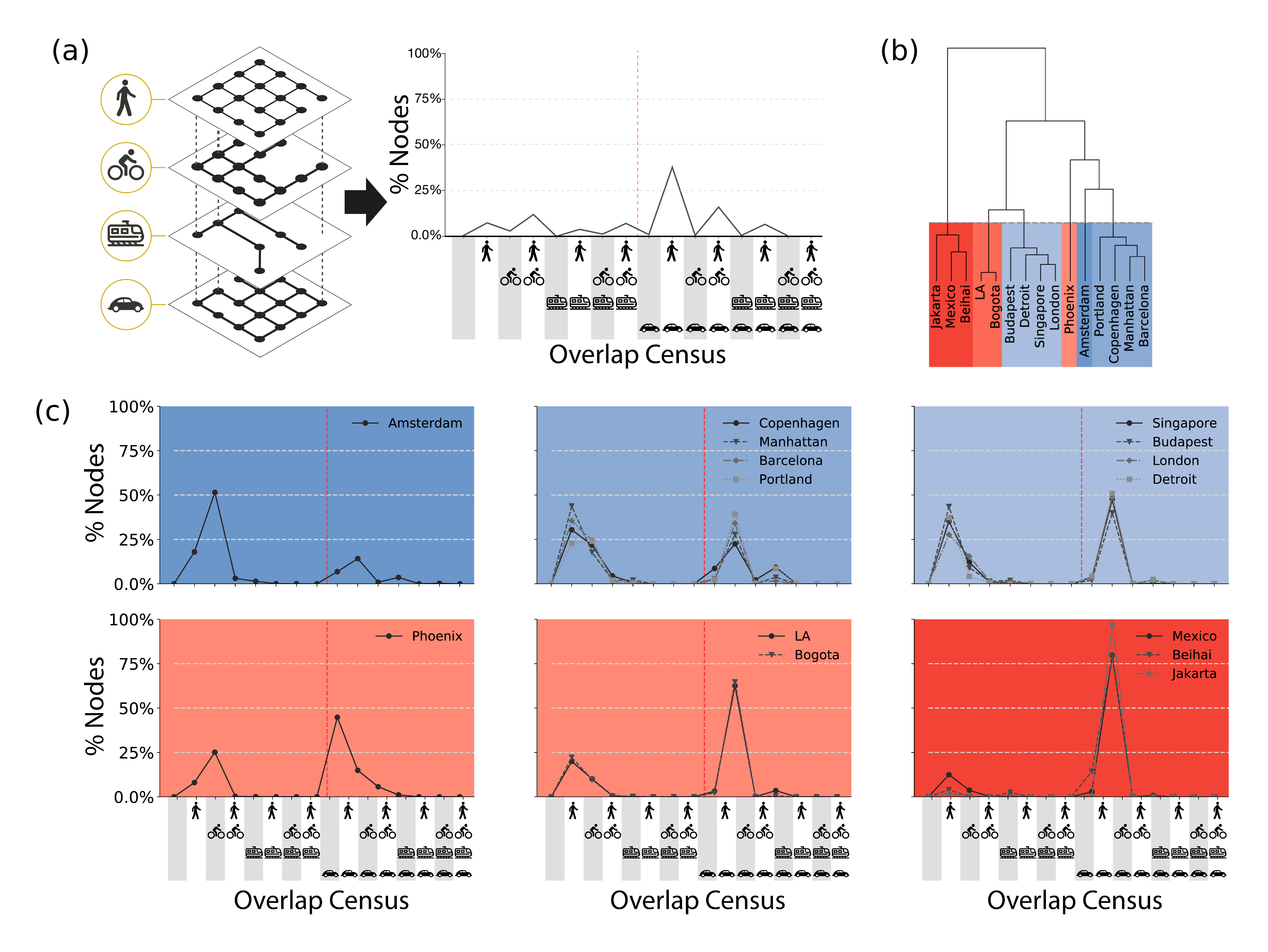}
	\caption{
		\textbf{(a)} Schematic of multiplex layers in a city (left) and its transformation to the overlap census (right). In the overlap census, the vertical red line gives a visual separation of the left from the right half where nodes become active in the street layer. High spikes in the right half indicate car-centricity.
		 \textbf{(b)} Clusters of cities based on similarity of their overlap census. We find six different clusters using a k-means algorithm (coloured areas), which explain more than $90\%$ of the variance.
		 \textbf{(c)} Overlap census for cities in each cluster. The first one corresponds to Amsterdam (the city with most active nodes in bicycle-only configurations). The Copenhagen-Manhattan-Barcelona-Portland city cluster has many active nodes in pedestrian-only and bicycle-only configurations, representing an active mobility city. The clusters of Los Angeles-Bogota and Mexico-Beihai-Jakarta are car-centric.}
	\label{fig:fig02}
\end{figure*}

Even in a multimodally ``optimal'' city there will be a high heterogeneity of node activities due to the different speeds and nature of transport modes, implying, for example, a much lower density of nodes necessary for a train network than for a bicycle network. Therefore, a good way to assess a city's overlap census is by comparing it with the overlap census of other cities. We find similarities between cities via a k-means algorithm fed with fifteen vectors (one per city), where each vector contains the percentages of nodes active in each possible configuration. The algorithm separates the 15 analyzed cities into six different clusters [Fig.~\ref{fig:fig02}(b)]. 

On the left half of the overlap census, we show the configurations in which nodes are not active in the street layer, while the right half contains car-related configurations [Fig.~\ref{fig:fig02}(c)] . These clusters of cities are useful to explain similarities in infrastructure planning in different transport development paths \citep{Rodrigue2013,Louf2014}, with clusters of car-centric urbanization (like Mexico, Beihai, and Jakarta) opposed to clusters that show a more multimodal focus in their mobility infrastructure (like Copenhagen, Manhattan, Barcelona, and Portland). In the extreme cluster that contains only Amsterdam, close to $50\%$ of nodes are active in the bicycle layer, whereas in the Mexico-Beihai-Jakarta cluster more than $50\%$ of nodes are active in the street-pedestrian configuration. The concentration of nodes in just one configuration informs not only about the mobility character of the city, i.e. Amsterdam being a bicycle-friendly city, but unveils the importance of explicitly considering overlooked layers and their interconnections. For example, Singapore, Budapest, London, and Detroit have two main peaks indicating that most of their nodes are either active in the street-pedestrian or only in the pedestrian configuration. This is not the case in Los Angeles and Bogota, where the majority of nodes are active in the car-pedestrian combination, i.e. the pedestrians have to share most of the city with cars. Our multimodal fingerprint unravels how different transport modes are interlaced, helping identifying which layer (or set of layers) could be improved to promote multimodal, sustainable mobility.

To summarize, we propose the new ``overlap census'' method based on multiplex network theory allowing to rigorously identify and compare the multimodal potential of cities.

\begin{small}
\setlength{\bibsep}{0.00cm plus 0.05cm} 
\bibliographystyle{apa}

\end{small}

\end{document}